\documentclass[pre,twocolumn,showpacs,superscriptaddress]{revtex4}
\usepackage{amsmath,amssymb,mathrsfs}
\usepackage[dvips]{color}
\usepackage{hyperref}
\usepackage{paralist}
\usepackage{epsfig}

\begin{document}

\title{Anomalous bulk behaviour in the free parafermion $Z(N)$ spin chain}

\author{Francisco C. Alcaraz}
\email[e-mail:]{alcaraz@ifsc.usp.br}
\affiliation{Instituto de F\'isica de S\~ao Carlos, Universidade de S\~ao Paulo,
Caixa Postal 369, 13560-970, S\~ao Carlos, SP, Brazil}

\author{Murray T. Batchelor}
\email[e-mail:]{batchelor@cqu.edu.cn}
\affiliation{Centre for Modern Physics, Chongqing University, Chongqing 400044, China}
\affiliation{Department of Theoretical Physics,
Research School of Physics and Engineering,
Australian National University, Canberra ACT 0200, Australia}
\affiliation{Mathematical Sciences Institute, Australian
National University, Canberra ACT 0200, Australia}

\begin{abstract}
We demonstrate using direct numerical diagonalization and extrapolation methods that boundary conditions have 
a profound effect on the bulk properties of a simple $Z(N)$ model for $N \ge 3$ for which the model hamiltonian is non-hermitian.
For $N=2$ the model reduces to the well known quantum Ising model in a transverse field. 
For open boundary conditions the $Z(N)$ model is known to be solved exactly in terms of free parafermions. 
Once the ends of the open chain are connected by considering the model on a ring, the bulk properties, including the 
ground-state energy per site, are seen to differ dramatically with increasing $N$. 
Other properties, such as the leading finite-size corrections to the ground-state energy, the mass gap exponent and the specific heat exponent, are also 
seen to be dependent on the boundary conditions. 
We speculate that this anomalous bulk behaviour is a topological effect.
\end{abstract}


\maketitle

\section{Introduction}

It is well known that non-hermitian systems are expected to behave differently to hermitian systems.
This is because non-hermitian hamiltonians describe the dynamics of physical systems that are not conservative.
Specifically, hermiticity guarantees that the energy spectrum is real and that time evolution is probability-preserving. 
Although there are many examples of integrable hermitian hamiltonians, integrable non-hermitian spin chain hamiltonians are relatively rare.
An important exception is the class of non-hermitian spin chains whose 
hamiltonians  are ${\cal P} {\cal T}$ symmetric, ensuring a real eigenspectrum \cite{Bender,others}.

Arguably the simplest of all exactly solved hermitian hamiltonians are those described by free fermions.
Indeed, the concept of free fermions plays an all pervasive and enduring role in the description of interacting classical and quantum spin systems.  
Recently it has become apparent that there is a simple exactly solved non-hermitian $Z(N)$ hamiltonian 
\begin{equation}
H_{\mathrm{open}}(L) = -  \sum_{j=1}^{L-1} \sigma_j \sigma_{j+1}^{\dagger} -  \lambda  \sum_{j=1}^L  \tau_j 
\label{ham}
\end{equation}
which displays the remarkable property of free parafermions \cite{Baxter1989a,Baxter1989b,Baxter2004,Fendley2014,ABL2017}, with 
a complex eigenspectrum.
This model is an $N$-state generalisation of the widely studied (hermitian) quantum Ising chain in a transverse field. 
Here $\sigma_j$ and $\tau_j$ are the usual $Z(N)$ operators, which in matrix form are defined by 
\begin{eqnarray}
\sigma_j &=& I \otimes I \otimes  \cdots \otimes I \otimes \sigma \otimes I \otimes \cdots \otimes I\\
\tau_j &=& I \otimes I \otimes  \cdots \otimes I \otimes \tau \otimes I \otimes \cdots \otimes I
\end{eqnarray}
where $I$, $\sigma$ and $\tau$ are each $N \times N$ matrices, with $\sigma$ and $\tau$ in position $j$. 
Here $I$ is the identity, with $\sigma$ and $\tau$ having components 
\begin{equation} 
\sigma_{m,n} = \omega^{m-1}\delta_{m,n}, \quad \tau_{m,n} = \delta_{m,n+1} 
\end{equation}
with $\omega = {\mathrm e}^{2\pi {\mathrm i}/N}$ and $\tau_{m,N}=\delta_{m,1}$. 
These are the clock  and shift matrices satisfying 
\begin{equation}
\sigma \tau = \omega \tau \sigma,  \quad 
\sigma^\dagger = \sigma^{N-1}, \quad \tau^\dagger = \tau^{N-1} 
\end{equation}
with $\sigma^N = \tau^N = I$. 
For $N=2$ they are the usual Pauli matrices $\sigma^z$ and $\sigma^x$.

The parameter $\lambda$ plays the role of temperature. 
Following \cite{ALC}, for the duality transformation for general $Z(N)$ quantum chains, it is 
simple to verify that hamiltonian (1) is self dual, namely $H(\lambda) = \lambda H(1/\lambda)$. 
We then expect, by usual arguments, that the model is critical at the self dual point $\lambda=\lambda_c =1$.
This is verified in the open boundary case, where the finite-size gaps are exactly known \cite{ABL2017}.

Generalizations of the hamiltonian (\ref{ham}) with the hermitian conjugate term included have been the subject of recent studies~\cite{hcon}, 
mostly for $N=3$, in the context of parafermionic edge modes~\cite{review}.
The unique property of hamiltonian (\ref{ham}) is that the energy eigenspectrum has the simple form 
\begin{equation}
-E/\lambda = \omega^{s_1} \epsilon_1 + \omega^{s_2} \epsilon_2 + \cdots + \omega^{s_L} \epsilon_L
\label{spec}
\end{equation}
for any choice of the integers $s_k = 0, \ldots,  N-1$.
This covers all $N^L$ eigenvalues in the spectrum.
Just as the fact that the special $N=2$ case $E/\lambda= \pm \epsilon_1 \pm  \epsilon_2 \pm \cdots \pm  \epsilon_L$ can be taken as the basic property of a 
free fermion system, the form (\ref{spec}) is the basic property of a free parafermion system.

The quasi energy levels $\epsilon_j$ $(j=1,\ldots,L)$ appearing in (\ref{spec}) are functions of $\lambda$.
Defining $g=1/\lambda^{N/2}$, 
the values $\epsilon_j^N$ are determined by the eigenvalues of the $L \times L$ matrices $C^\dagger C$ or $C C^\dagger$, where 
\begin{equation}
C=\begin{bmatrix} 1 &  &  & &\\ g &1&&&\\  & g & 1 & &\\  &  & \ddots & \ddots & \\& &  & g & 1 
\end{bmatrix}
\end{equation}
with
\begin{equation}
\epsilon_{j} = \left( 1 + g^2 + 2 g \cos {k_j} \right)^{1/N}.
\end{equation}
The roots $k_j$, $j=1,\ldots,L$, satisfy the equation \cite{ABL2017} 
\begin{equation}
{\sin(L+1) k = - g \sin Lk}.
\label{eqn}
\end{equation}
Using this solution a number of exact results have been derived for this model \cite{ABL2017}. 
Although having a simpler hamiltonian than the free fermionic superintegrable chiral Potts model, 
the free parafermionic model is seen to share some critical properties with it, namely 
the specific heat exponent $\alpha=1-2/N$ and 
the anisotropic correlation length exponents $\nu_\parallel =1$ and $\nu_\perp=2/N$.

Here we consider the more general hamiltonian 
\begin{equation}
H(L,a) = H_{\mathrm{open}}(L) - a \, \sigma_L \sigma_1^\dagger 
\label{gen}
\end{equation}
where $H_{\mathrm{open}}(L)$ is as defined in (\ref{ham}) and $a$ is a 
{real} parameter interpolating between periodic boundary conditions (PBC) ($a=1$) and anti-periodic boundary conditions ($a=-1$). 
Obviously $a=0$ recovers the model with open boundary conditions (OBC). 
The motivation for the present study is to investigate the role of boundary conditions on the properties of the free parafermion $Z(N)$ model for 
$N \ge 3$ \cite{footnote}. 
As discussed for the chiral Potts model 
from the perspective of conformal field theory \cite{Cardy1993}, 
several of the usual properties of hermitian systems, 
such as insensitivity of bulk thermodynamic quantities to boundary conditions, 
can fail in the non-hermitian case.
As foreshadowed, this note of caution applies even more so for the model under consideration \cite{ABL2017}.
We report here that the role of boundary conditions is seen to have a profound effect on the bulk properties of the 
non-hermitian free parafermion $Z(N)$ hamiltonian.

\section{Bulk ground state energy per site}

\subsection{Periodic boundary conditions}

As remarked above, the $Z(N)$ model defined in Eq.~(\ref{gen}) is solved exactly  for general $N$ and finite $L$ for the case of OBC ($a=0$). 
For PBC ($a=1$) we resort to numerical diagonalization to 
calculate the ground-state energy per site $e_L=E_0(L)/L$ 
for the $Z(N)$ model for chain sizes $L=2,3,\ldots,L_{\mathrm{max}}$. 
For comparison we also consider OBC in the same way.
From the energy expression (\ref{spec}) it is evident that the ground-state energy is real for OBC, corresponding to the integers $s_k = 0$ for all $k$. 
For PBC, although no similar such exact solution has been obtained for PBC, 
we observe that the ground-state energy is also real.
A proof of this observation, based on symmetries of these quantum chains is still missing.

\begin{figure}[h]
\includegraphics[angle=0,width=0.45\textwidth] {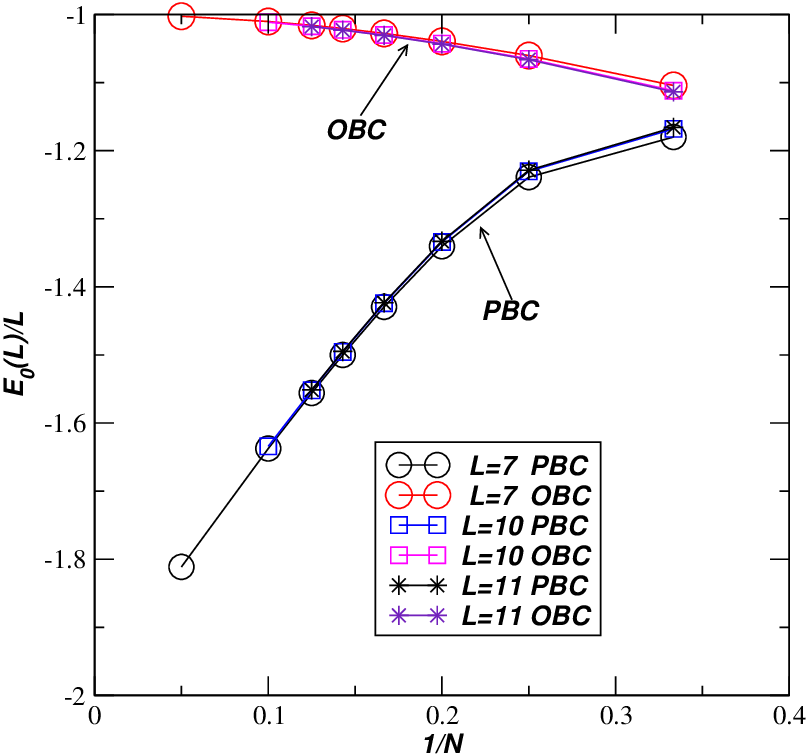}
\caption{
The ground-state energy per site for the $Z(N)$ spin chain with periodic boundary conditions (PBC) and open boundary conditions (OBC) 
for $N=3,4,5,6,7,8,10$ and $20$. The data points (see legend) are the values for the $Z(N)$ model  
for chain sizes $L=7$, $L=10$ and $L=11$.}
\label{figurea}
\end{figure}

\begin{figure}[h]
\includegraphics[angle=0,width=0.45\textwidth] {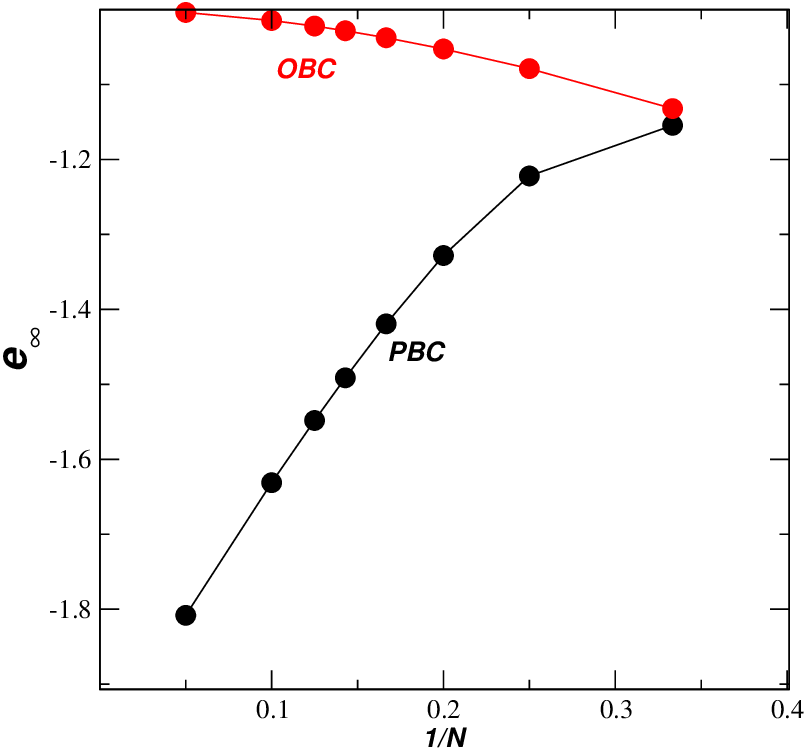}
\caption{Depiction of the contrast between  
the extrapolated estimates for the ground-state energy per site for the $Z(N)$ model with periodic boundary conditions (PBC) and 
open boundary conditions (OBC) for $N=3,4,5,6,7,8,10$ and $20$. 
These results are the values shown in Table~\ref{table1}.}
\label{figureb}
\end{figure}

In the present study, we concentrate on the value $\lambda=1$. 
The values for the ground-state energy per site are plotted 
for some fixed chain sizes and different values of $N$ in Fig.~\ref{figurea}.
%
We clearly see that for a given size $L$, the difference between $e_L$ for PBC and OBC increases with $N$. 
Moreover, while $e_L$ increases with $N$ for OBC, it decreases with $N$ for PBC.
Extrapolated estimates for $e_\infty$ are shown in Table~\ref{table1}.
The extrapolations were performed using van der Broeck-Schwartz extrapolants with $\epsilon$-extension (VBS) \cite{VBS}. 
In each case the error indicated is an evaluation taking into account the stability as $\epsilon$ is changed in the extrapolation.
The estimates for $e_\infty$ are visualized in Fig.~\ref{figureb}, 
which shows the striking dependence of the bulk ground-state energy per site on the boundary conditions.
The known exact result for $e_\infty$ with OBC is given further below in Eq.~(\ref{eq-3-1}), with $e_\infty=-1$ in the limit $1/N \to 0$.

\begin{table*}[t]
\begin{center}
\begin{tabular}{lcccc}
\cline{1-5}

           & $L_{\mbox{max}}$ & Extrap. PBC  & Extrap. OBC & Exact OBC  \\ \hline
\hline
$Z(3)$ & 21 & $-1.1544 \pm 0.0002$ & $-1.1321 \pm 0.0002$ & $-1.13209336...$  \\ \hline
$Z(4)$ & 17 & $-1.2219 \pm 0.0002$ & $-1.0787 \pm 0.0001$ & $-1.07870520...$  \\ \hline 
$Z(5)$ & 14 & $-1.3280 \pm 0.0002$ & $-1.0524 \pm 0.0001$ & $-1.05246524...$  \\ \hline 
$Z(6)$ & 13 & $-1.4192 \pm 0.0002$ & $-1.0375 \pm 0.0001$ & $-1.03754819...$ \\ \hline 
$Z(7)$ & 12 & $-1.4913 \pm 0.0002$ & $-1.0282 \pm 0.0001$ & $-1.02823144...$  \\ \hline 
$Z(8)$ & 11 & $-1.5482 \pm 0.0001$ & $-1.0220 \pm 0.0001$ & $-1.02201332...$  \\ \hline 
$Z(10)$& 10 & $-1.6312 \pm 0.0002$ & $-1.0145 \pm 0.0001$ & $-1.01447454...$  \\ \hline 
$Z(20)$& 7  & $-1.8080 \pm 0.0004$ & $-1.0038 \pm 0.0001$ & $-1.00384106...$  \\ \hline 
\end{tabular}
\end{center}
\caption{Estimated results for the ground-state energy per site of the $Z(N)$ model  
for periodic boundary conditions (PBC) and for open boundary conditions (OBC). 
The extrapolated values are obtained for chain sizes $L=2,3,\ldots,L_{\mathrm{max}}$. 
The exact values for OBC are also shown.}
\label{table1}
\end{table*}

\subsection{General boundary conditions}

\begin{figure}
\centering
\includegraphics[angle=0,width=0.45\textwidth] {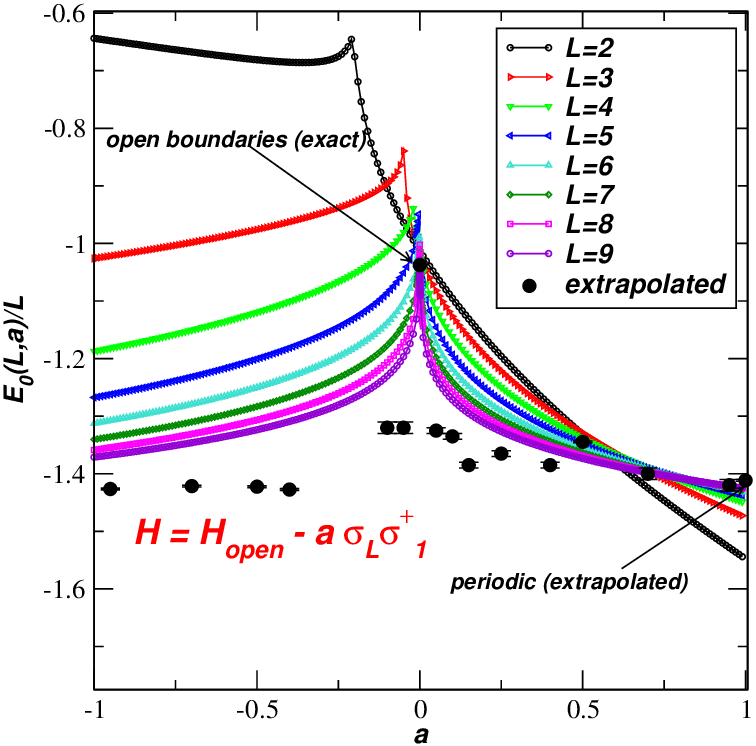}
\caption{
The ground-state energy per site for the $Z(6)$ model (\ref{gen}) for the general boundary conditions 
defined by the parameter $a$. The VBS-extrapolated results are also shown (the deviations 
 in the extrapolations are subjective).}
\label{fig3}
\end{figure}

\begin{figure}
\centering
\includegraphics[angle=0,width=0.45\textwidth] {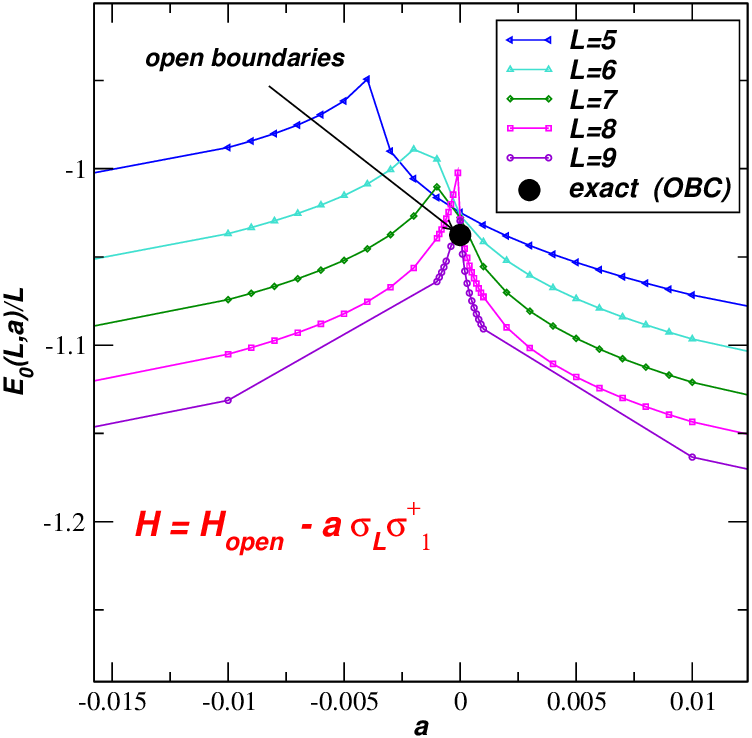}
\caption{
The ground-state energy per site for the $Z(6)$ model (\ref{gen}) for the general boundary conditions 
defined by the parameter $a$. The exact value for the open boundary case is shown.}
\label{fig4}
\end{figure}

In order to further investigate the effect of the boundary conditions, we now consider the 
general boundary hamiltonian $H(L,a)$ given in Eq.~(\ref{gen}).
Here the parameter $a$ interpolates between the open and periodic cases. 
In Fig.~\ref{fig3} we show the values of $e_L(a)=E_0(L,a)/L$ for the $Z(6)$ model for chain sizes $L=2-9$. 
We see in this figure the existence of peaks as a function of the parameter $a$. 
As $L$ becomes larger the peaks tend to the position $a=0$, i.e., the OBC case, and become sharper as the chain size grows. 
In Fig.~\ref{fig4} we show the curves of Fig.~\ref{fig3} in a larger scale around $a=0$, at which the exact result is known. 
These figures appear to indicate that, except for the OBC $a=0$, all the closed boundaries $a\neq 0$ 
have the same value for the ground-state energy per site in the infinite size limit. 
In Fig.~\ref{fig3} we also show the values obtained from the VBS-extrapolations using the lattice sizes $L=2-9$.
Here the errors shown in the extrapolation are not errors in the strict sense, but rather subjective evaluations taking into account the 
behavior of the extrapolations.

\begin{figure}
\centering
\includegraphics[angle=0,width=0.45\textwidth] {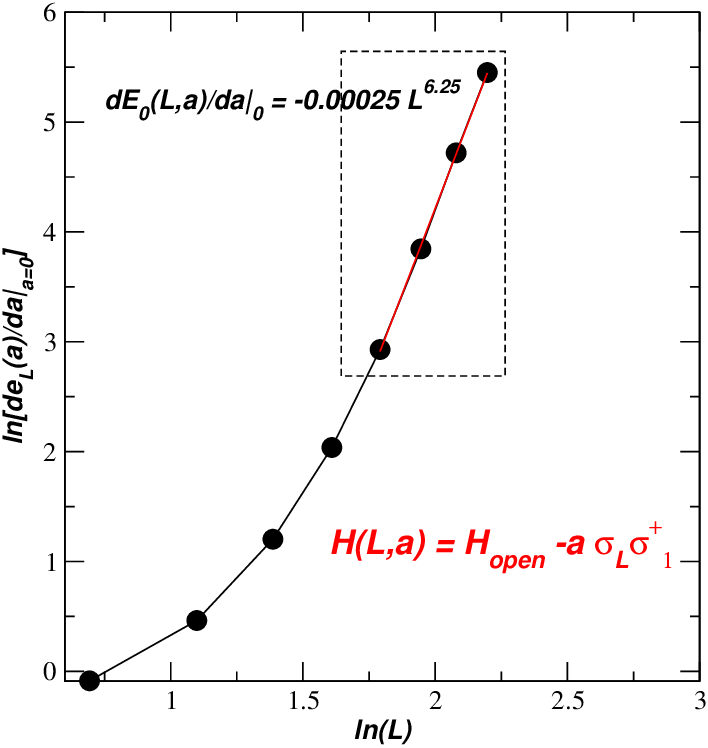}
\caption{
A log-log plot of the derivative $de_L(a)/da |_{a=0}$ as a function of $1/L$ for the $Z(6)$ model (\ref{gen}) with boundaries specified by the parameter $a$.}
\label{fig5}
\end{figure}

In order to confirm the abnormal behavior at $a=0$ we compute numerically the derivative  $e'_L(L,a)=de_L(a)/da |_{a=0}$. 
Specifically, we compute the right-derivative
\begin{eqnarray} \label{eq4-4}
 \frac{df(x)}{dx} &=& \frac{-3f(x) +4 f(x+\Delta x) -f(x+2\Delta x)}{2\Delta x} \nonumber \\
 && + \, O((\Delta x)^2).  
\end{eqnarray}
The results for this derivative up to $L=9$ are shown in Table II for the $Z(6)$ model.
These values are shown in a log-log plot in Fig.~\ref{fig5}. 
We clearly see that the derivatives diverge to $-\infty$ polynomially with $L$. 
A fit for the $Z(6)$ model, obtained from the chain sizes $L=6-9$ (dashed 
rectangle in Fig.~5), gives $de_L(a)/da |_{a=0} \approx -0.00025 L^{6.25}$. 
The tendency for an infinite derivative can also be seen in Fig.~6, 
where we plot the inverse of the derivative as a function of $1/L$. 
Here the tendency is clearly towards the value zero as $L\to \infty$.

\begin{table*}[htp]
\begin{center}
\begin{tabular}{lr}
\cline{1-2}
   $L$        & $e'_L(L,a) |_{a=0}$ \\ 
\hline
\hline
          2 & $-0.91763825$ \\ \hline     
          3 & $-1.58769897$ \\ \hline    
          4 & $-3.32838276$ \\ \hline    
          5 & $-7.67373638$ \\ \hline    
          6 & $-18.7154986$ \\ \hline    
          7 & $-46.7908356$ \\ \hline    
          8 & $-112.234431$ \\ \hline    
          9 & $-233.157167$ \\ \hline    
\end{tabular}
\end{center}
\caption{The derivative $de_L(a)/da |_{a=0}$ with increasing chain size $L$ for the $Z(6)$ model (\ref{gen}) with boundary condition specified by the parameter $a$.}
\label{table2}
\end{table*}

\begin{figure}
\centering
\includegraphics[angle=0,width=0.45\textwidth] {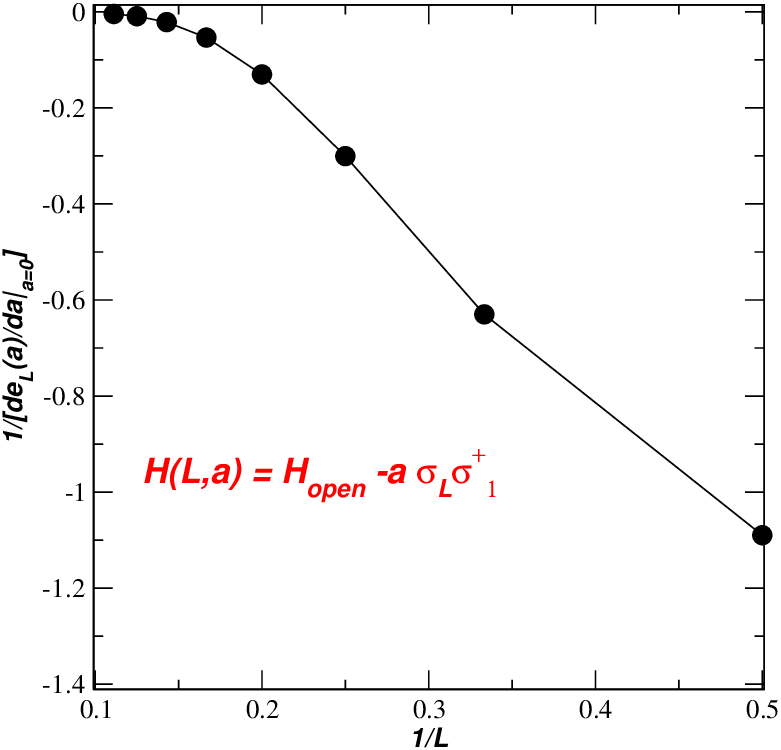}
\caption{
The inverse of  the derivative $de_L(a)/da |_{a=0}$ as a function of $1/L$ for the $Z(6)$ model (\ref{gen}) with 
boundary conditions specified by the parameter $a$.}
\label{fig6}
\end{figure}

\subsection{Leading finite-size corrections}

\begin{figure}
\vskip 7mm
\centering
\includegraphics[angle=0,width=0.45\textwidth] {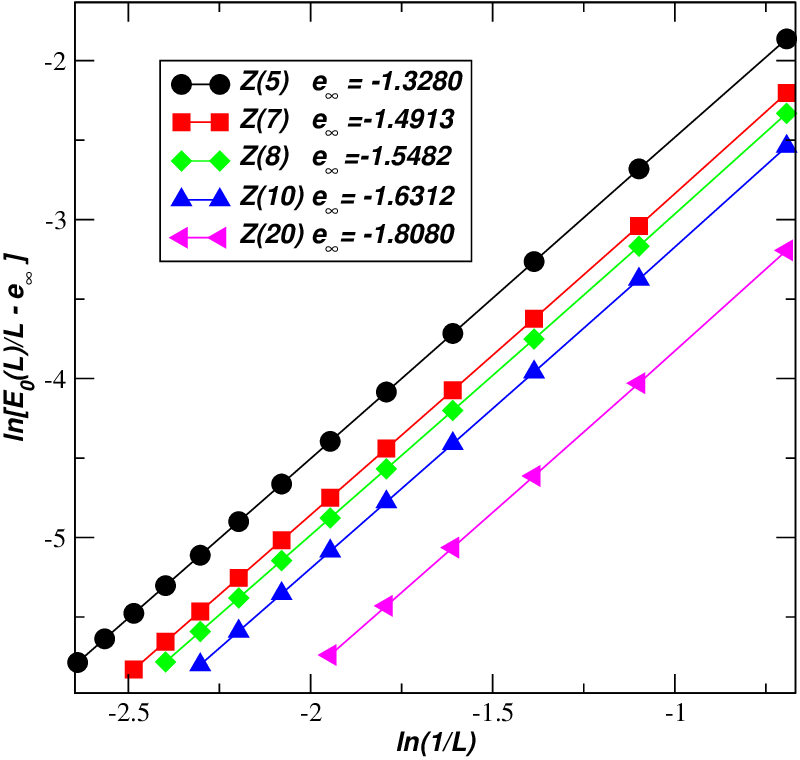}
\caption{
The fittings, following Eq. (\ref{eq-fit}), for the ground-state energy per site $E_0(L)/L$ 
as a function of $1/L$ for the $Z(N)$ model ($N=5,7,8,10$ and $20$) with periodic 
boundaries. The values $e_{\infty}$ in the bulk limit are shown in the inset.}
\label{fig7}
\end{figure}

In the open boundary case the leading finite-size corrections to the ground-state energy are known to be given exactly by \cite{ABL2017}
\begin{equation} \label{eq3}
 E_0(L) = Le_{\infty} + f_{\infty} + \frac{b_N}{L^{\nu}} + O(\frac{1}{L^{1+\nu}})
\end{equation}
where 
\begin{equation} \label{eq-3-1}
e_{\infty} = -\frac{2^{\nu}}{\sqrt{\pi}}\frac{\Gamma(\frac{1}{2}+\frac{1}{N})}{\Gamma(1+\frac{1}{N})}, \quad f_{\infty} = \frac{1}{2}e_{\infty} + 2^{\nu -1} 
\end{equation}
and $\nu = 2/N$. 
The amplitude $b_N$ is also known. 
In the periodic case we would expect the leading behavior to be of the form 
\begin{equation} \label{eq-fit}
\frac{E_0(L)}{L} = e_{\infty} + \frac{b}{L^{\gamma}} +o(1/L^{\gamma})
\end{equation}
with the exponent value $\gamma = 1+ \nu$. 
To test this we have evaluated the exponent $\gamma$ in two distinct ways. 
Firstly we have made a fit where $e_{\infty}$, $b$ and $\gamma$ are free parameters.  
Secondly we take the extrapolated values shown in Table~\ref{table1} for the ground-state energy per site $e_{\infty}$ and then perform a fit of the form
\begin{equation} \label{eq4}
\frac{E_0(L)}{L} - e_{\infty} = \frac{b}{L^{\gamma}} 
\end{equation}
with $b$ and $\gamma$ taken as free parameters. 
For the sake of illustration we show in Fig.~\ref{fig7} the various fittings for the $Z(N)$ model for values $N=5, 7, 8, 10$ and 20.

The values obtained by the two procedures are shown in Table~\ref{table3}. 
In columns 2-4  of Table~\ref{table3} we show the results obtained for the exponent via the first method, 
with the results obtained via the second method shown in column 5.
We believe that the second method is more reliable since 
it takes into account the extrapolated values of $e_{\infty}$, given in Table~\ref{table1}. 
Taking into account both methods we give the estimate shown in column 6, 
where the error is an indication of the expected precision (clearly subjective).

\begin{table*}[htp]
\begin{center}
\begin{tabular}{ccccccc}
\cline{1-7}
   $N$        & $e_{\infty} \mbox{(fit)}$ & $b \mbox{(fit)}$ & 
$\gamma \mbox{(fit)}$ & $\gamma \mbox{(extr)}$ & $ \gamma$ & $\gamma_{\mathrm{open}}$ \\ \hline
\hline
3 & $-1.15355$ & $-0.68$ & 1.68 & 1.70 & 1.68 $\pm$ 0.02 & 1.67 \\ \hline     
4 & $-1.22118$ & $-0.72$ & 1.89 & 1.92 & 1.90 $\pm$ 0.03 & 1.50 \\ \hline     
5 & $-1.32810$ & $-0.63$ & 2.02 & 2.02 & 2.02 $\pm$ 0.02 & 1.40 \\ \hline     
6 & $-1.41952$ & $-0.53$ & 2.05 & 2.01 & 2.03 $\pm$ 0.03 & 1.33 \\ \hline     
7 & $-1.49135$ & $-0.46$ & 2.06 & 2.02 & 2.04 $\pm$ 0.03 & 1.29 \\ \hline     
8 & $-1.54849$ & $-0.40$ & 2.06 & 2.03 & 2.04 $\pm$ 0.03 & 1.25 \\ \hline     
10& $-1.63144$ & $-0.33$ & 2.06 & 2.02 & 2.04 $\pm$ 0.03 & 1.20  \\ \hline     
20& $-1.80820$ & $-0.17$ & 2.07 & 2.03 & 2.05 $\pm$ 0.03 & 1.10  \\ \hline     
\end{tabular}
\end{center}
\caption{The PBC values $e_{\infty}(\mbox{fit})$, $b(\mbox{fit})$ and $\gamma(\mbox{fit})$ 
are the results obtained by fitting the finite-size correction form given in Eq.~(\ref{eq-fit}). 
The PBC values $\gamma \mbox{(extr)}$ are obtained using the extrapolated values for $e_{\infty}$ in Eq.~(\ref{eq4}). 
The second last column shows the estimated values for the PBC finite-size correction exponent $\gamma$ taking into account both methods. 
Also shown for comparison is the exponent $\gamma_{\mathrm{open}}$ obtained by using the same lattice sizes in the extrapolation.}
\label{table3}
\end{table*}

We clearly see from the results of Table~\ref{table3} that the leading finite-size correction for the ground-state energy 
is governed by the exponent values $\gamma \approx 2$ for $N \geq 4$, which are   
quite distinct from the corresponding values with OBC, namely $\gamma = 1 + 2/N$. 
For comparison of the methods, we also show, up to two decimal digits, the values 
obtained in this way for the exponent $\gamma$ in the OBC case, using the same lattice sizes as in the periodic case. 
They are in close agreement with the known result.

\section{Gap exponent}

The excitation energies above the ground-state, and consequently the energy gaps of the parafermionic models 
have  complex values, irrespective of whether the boundary conditions are open or periodic. 
Although some energy levels are real, those with lowest real part are complex. 
In this section we consider the gap with lowest real part. 
The model (\ref{gen}) has a $Z(N)$ symmetry, due to the commutation relation
\begin{equation} \label{eq8}
[H,{\cal{P}}] =0, \quad {\cal{P}}=\prod_{j=1}^L \tau_j.
\end{equation}
The ground-state belongs to the $Z(N)$ charge ${\cal P} =0$, with the first gap to 
the sector of charge ${\cal P}=1$. 
The correlation length exponent $\nu$ can be estimated from the leading finite-size behavior of the first gap, with
\begin{equation} \label{eq9}
 G_L = \mbox{Re}\{ E_1(L) - E_0(L)\}=\frac{A}{L^{\nu}} +o(1/L^{\nu})
\end{equation}
where $A$ is a constant. 
We consider the finite-size estimator for the exponent $\nu$ defined by
\begin{equation} \label{eq9p}
 \nu_{L,L+1} = \frac{ \ln (G_L/G_{L+1})}{\ln((L+1)/L)}.
\end{equation}

\begin{figure}
\centering
\includegraphics[angle=0,width=0.45\textwidth] {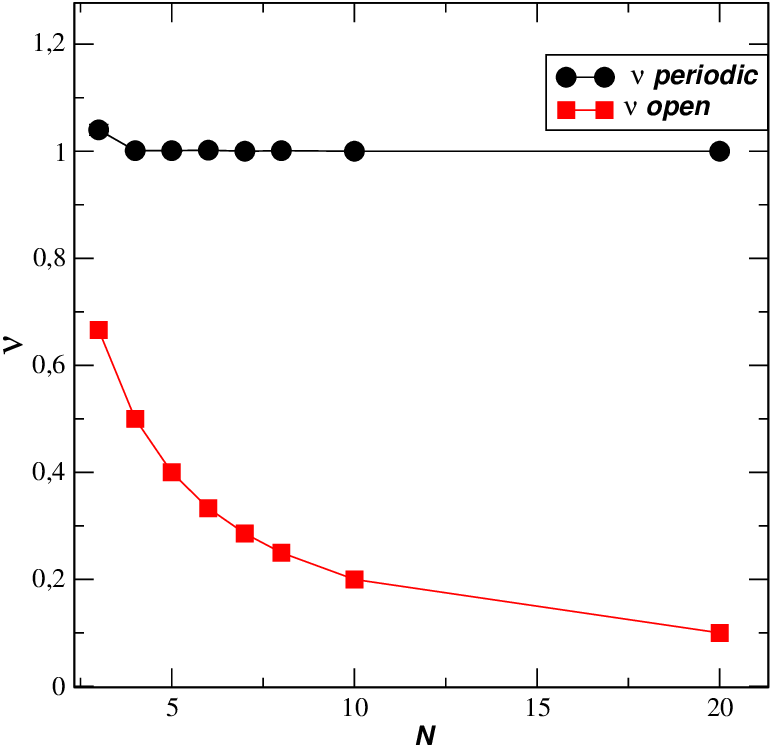}
\caption{
The results for the exponent $\nu$ obtained from the VBS extrapolations of 
the estimator (\ref{eq9p}) for the periodic $Z(N)$ model 
and the corresponding exact results for OBC.}
\label{fig8}
\end{figure}

In Table~\ref{table4} we show the results obtained from VBS-extrapolants of the data for $\nu_{L,L+1}$. 
We show in the third column the results with our subjective evaluation of the errors. 
We also show in this table the results obtained for the exponents for OBC, using the same chain sizes. 
In the last column we show the known exact results for OBC.
We clearly see that the values of the gap exponent $\nu$ are quite distinct for PBC vs OBC.   
It seems that the exponent for the periodic case is close to (if not exactly) the value $\nu=1$, in distinction to OBC where $\nu=2/N$. 
To illustrate this difference we show in Fig.~\ref{fig8} the extrapolated results for PBC together with the exact results for OBC.

\begin{table*}[htp]
\begin{center}
\begin{tabular}{lcccl}
\cline{1-5}
   $N$        & $\nu$(extr.) & $\nu$ (predicted) & $\nu_{\mathrm{open}}$(extr.) & $\nu_{\mathrm{open}}$(exact) \\ 
\hline
3 & 1.080   & $1.080  \pm 0.005$  & 0.667        & $2/3=0.666\ldots$   \\ \hline     
4 & 1.005   & $1.005  \pm 0.003$  & 0.500        & $2/4=0.5$           \\ \hline     
5 & 1.001   & $1.001 \pm 0.002$   & 0.400        & $2/5=0.4$           \\ \hline     
6 & 1.002   & $1.002 \pm 0.002$   & 0.333        & $2/6=0.333\ldots$   \\ \hline     
7 & 1.000   & $1.000 \pm 0.001$   & 0.288        & $2/7=0.2857\ldots$  \\ \hline     
8 & 1.000   & $1.000 \pm 0.001$   & 0.250        & $2/8=0.25$          \\ \hline     
10 & 1.000   & $1.000 \pm 0.001$   & 0.200        & $2/10=0.2$          \\ \hline     
20 & 1.000   & $1.000 \pm 0.001$   & 0.100        & $1/10=0.1$          \\ \hline     
\end{tabular}
\end{center}
\caption{The gap exponent $\nu$ obtained for the periodic $Z(N)$ model using 
the VBS extrapolation of the estimators (\ref{eq9p}). Also shown are the  
results obtained for OBC with the same chain sizes used in the periodic case. 
The exact results for OBC are shown in the last column.}
\label{table4}
\end{table*}

\section{Specific heat exponent}

We calculate in this section the specific heat of the $Z(N)$ model with PBC  
at the critical point $\lambda =\lambda_c =1$. 
This quantity is given by
\begin{equation} \label{eq14}
C(\lambda,L) = -\frac{1}{L} \frac{d^2E_0(L)}{d\lambda}.
\end{equation}
At the critical point we should expect the leading finite-size behavior
\begin{equation} \label{eq15}
C(\lambda=1,L) \sim A \, L^{\alpha/\nu_{\parallel}}
\end{equation}
where $A$ is a constant. 
In the case of OBC, $\alpha = 1-2/N$ and $\nu_{\parallel}=1$ \cite{ABL2017}. 
In the periodic case the finite-size values of (\ref{eq15}) are given in Table~\ref{table5} for the $Z(N)$ model with $N=3,5,6,7$ and 8. 
Surprisingly, we see that the data saturates as $L$ increases with a clear indication that the specific heat exponent $\alpha =0$ for the periodic case, 
as for the $N=2$ Ising model.
Actually the results we have obtained show that the periodic case, at least for $N>4$ exhibits a similar  
 behavior as the standard Ising model. 
This fact should be explored further  in subsequent studies. 

\begin{table*}[htp]
\begin{center}
\begin{tabular}{lccccc}
\cline{1-6}
 $L$ &  $N=3$        & $N=5$ & $N=6$ & $N=7$ & $N=8$ \\ 
\hline
    2 &   0.433013 &   0.248680 &   0.175466 &   0.117594 &   0.092118\\
    3 &   0.629961 &   0.278889 &   0.189414 &   0.130737 &   0.105481\\
    4 &   0.755042 &   0.278853 &   0.191451 &   0.135007 &   0.110214\\
    5 &   0.840759 &   0.276337 &   0.192507 &   0.137145 &   0.112457\\
    6 &   0.901140 &   0.274801 &   0.193252 &   0.138354 &   0.113684\\
    7 &   0.943967 &   0.274056 &   0.193770 &   0.139095 &   0.114426\\
    8 &   0.974148 &   0.273712 &   0.194129 &   0.139580 &   0.114908\\
    9 &   0.995022 &   0.273552 &   0.194384 &   0.139914 &   0.115238\\
   10 &   1.008975 &   0.273475 &   0.194570 &   0.140154 &   0.115475\\
   11 &   1.017767 &   0.273437 &   0.194710 &   0.140331 &   0.115650\\
   12 &   1.022719 &   0.273417 &   0.194816 &   0.140466 &      -   \\
   13 &   1.024835 &   0.273406 &      -    &      -    &      -   \\
   14 &   1.024883 &   0.273401 &      -    &      -    &      -   \\
   15 &   1.023453 &      -    &      -    &      -    &      -   \\
   16 &   1.020994 &      -    &      -    &      -    &      -   \\
   17 &   1.017848 &      -    &      -    &      -    &      -   \\
   18 &   1.014273 &      -    &      -    &      -    &      -   \\
   19 &   1.010465 &      -    &      -    &      -    &      -   \\
   20 &   1.006565 &      -    &      -    &      -    &      -   \\
\end{tabular}
\end{center}
\caption{The specific heat $C(\lambda=1,L)$ for the $Z(N)$ model with $L$ sites, 
for $N=3,5,6,7$ and 8.} 
\label{table5}
\end{table*}

\section{Summary and Discussion}

The bulk properties of the $Z(N)$ model defined by the non-hermitian hamiltonian (\ref{gen}) have been demonstrated here to exhibit  
a striking dependence on boundary conditions.
For illustrative purposes we have focussed on the critical point $\lambda=1$.
For $N=2$, the widely studied hermitian quantum Ising chain in a transverse field, the bulk properties are well known to be independent of the boundary conditions. 
As can be seen clearly in Fig.~\ref{figureb},   
the difference between the values obtained for the bulk ground-state energy per site $e_\infty$ with OBC ($a=0$) and PBC ($a=1$) increases with increasing $N$ for $N\ge 3$. 
As a function of the boundary condition parameter $a$, the bulk ground-state energy per site 
is a singular point at $a=0$, as can be seen for the $Z(6)$ model in Fig.~\ref{fig3} and Fig.~\ref{fig4}.
We observed the divergence of the derivative with respect to the parameter $a$ at $a=0$.
This is precisely the open boundary case.

The finite-size corrections to the bulk ground-state energy per site are also dependent on the boundary conditions.
We found that for PBC the leading finite-size correction to the bulk ground-state energy is of the form (\ref{eq-fit}) 
governed by the exponent values $\gamma \approx 2$ for $N \geq 4$, which are   
distinct from the corresponding exactly known values for OBC, namely $\gamma = 1 + 2/N$.

The first mass gap exponent has also been numerically estimated for PBC, with values  for all $N$ close to the Ising $N=2$  value $\nu=1$.
This result is again strikingly different to the known value $\nu=2/N$ for OBC, recall Fig.~\ref{fig8}.
Moreover, the analysis of the specific heat in Section IV indicates that for PBC the values of the specific heat exponent $\alpha$ 
are also suggestive, at least for $N>4$, of the Ising model value $\alpha=0$.
The fact that for the periodic case, for large $N$, the exponent $\gamma$ in (\ref{eq4}) is close to 2 suggests we have a relativistic energy-momentum 
dispersion relation, and possibly an underlying conformal invariance 
in the bulk limit.
Since for large $N$ the exponents $\nu\approx 1$ and $\alpha\approx 0$, the natural possibility would be the Ising universality class with  
central charge $c=1/2$. In order to test this possibility we have calculated 
the mass gaps with lowest real part in the eigensectors labeled by
the momentum ${2\pi} p/{L}$ ($p=0,1,\ldots,N-1$) and $Z(N)$ charges 
($Q=0,\ldots,N-1$) of the $Z(8)$ quantum chain with $L=10$. Exploring the 
well  known consequences of conformal invariance,  the 
mass gap amplitudes of finite lattices give us  predictions for the 
conformal dimensions in clear contradiction with the 
expected results of  an Ising conformal field theory. 
%

At this stage we can only begin to speculate on the reasons for why the boundary conditions have such a profound effect on the bulk properties of this simple $Z(N)$ model. 
Systems for which the boundary conditions affect the finite-size corrections are usual, normally producing an additional surface term of $O(1/L)$ in the energy. 
There also exist systems where the mass gap and critical behavior may change or even vanish under change of boundary conditions. 
An example is the non-hermitian hamiltonian associated with the time-evolution operator of the asymmetric exclusion process  
where the open problem is gapped (the hamiltonian is related to the XXZ quantum chain in the gapped ferromagnetic regime), 
but the closed system is gapless and critical (in the KPZ universality class) \cite{ASEP1,ASEP2,AR1}. 
However, the ground-state energies (with value zero in this example) are the same for both boundary conditions. 
Systems for which the bulk energy changes with the boundary conditions are surprising exceptions.
A prominent example for two-dimensional classical systems is the six-vertex model with domain wall boundary conditions, 
for which the bulk free energy differs from the well known result obtained using periodic or open boundary conditions~\cite{Korepin}. 
For the model under consideration here it took some time for us to be fully convinced by our numerical results. 
For the periodic $Z(N)$ model at $\lambda=1$ the ground-state energy per site 
decreases with increasing $N$, in contrast to the open case where it increases.
The ordinary $Z(N)$ hermitian quantum chains like the Potts or the 
$Z(N)$ parafermionic models \cite{FZ} give a bulk ground-state energy which 
is independent of the boundary conditions and decreases with increasing $N$ \cite{FCA1,FCA2}.
This suggests that the ground-state of the $Z(N)$ model with open ends is constrained (probably topologically restricted), 
but by insertion of a single link connecting both sides of the chain, and thereby changing the lattice topology, 
the energy of the ground-state is decreased enormously (by $O(L)$).
Conversely, the physics of the $Z(N)$ model defined on a ring changes drastically by cutting a single link.
In this sense it is the $Z(N)$ model with OBC which is the exceptional case.

Here we can also throw into the mix the fact that the $Z(N)$ model with OBC is described by the physics of free parafermions.
The free parafermion description works perfectly for this model when subject to OBC, 
but there is of course no guarantee of a solution  in terms of free parafermions for PBC. 
The underlying reason may thus again be topological and related to the ordering of the parafermionic operators.

\vskip 1cm

{\em Acknowledgments.} 
The work of FCA is supported in part by the Brazilian agencies FAPESP and CNPq. 
The work of MTB is supported by The 1000 Talent Program of China,  
National Natural Science Foundation of China Grant No.~11574405 
and Australian Research Council Discovery Project DP180101040.

\end{document}